\documentclass[twocolumn]{aastex6}
\usepackage{times}
\usepackage{amsmath}
\usepackage{textcomp}
\usepackage{amssymb}

\newcommand{\lum}{erg\,s$^{-1}$}
\newcommand{\fermi}{{\it Fermi}}

\newcommand{\swift}{{\it Swift}}
\newcommand{\phflux}{\mbox{${\rm \, ph \,\, cm^{-2} \, s^{-1}}$}}

\newcommand{\gm}{$\gamma$}

\shorttitle{High redshift {\it Fermi} Blazars}

\begin{document}

\title{Gamma-ray blazars within the first 2 billion years}
\author{
M.~Ackermann\altaffilmark{2}, 
M.~Ajello\altaffilmark{3,1}, 
L.~Baldini\altaffilmark{4}, 
J.~Ballet\altaffilmark{5}, 
G.~Barbiellini\altaffilmark{6,7}, 
D.~Bastieri\altaffilmark{8,9}, 
J.~Becerra~Gonzalez\altaffilmark{10,11}, 
R.~Bellazzini\altaffilmark{12}, 
E.~Bissaldi\altaffilmark{13}, 
R.~D.~Blandford\altaffilmark{14}
% ...
}
\altaffiltext{1}{Corresponding authors: M.~Ajello, majello@g.clemson.edu; D.~Gasparrini, gasparrini@asdc.asi.it; R.~Ojha, roopesh.ojha@gmail.com; V.~S.~Paliya, vpaliya@g.clemson.edu.}
\altaffiltext{2}{Deutsches Elektronen Synchrotron DESY, D-15738 Zeuthen, Germany}
\altaffiltext{3}{Department of Physics and Astronomy, Clemson University, Kinard Lab of Physics, Clemson, SC 29634-0978, USA}
\altaffiltext{4}{Universit\`a di Pisa and Istituto Nazionale di Fisica Nucleare, Sezione di Pisa I-56127 Pisa, Italy}
\altaffiltext{5}{Laboratoire AIM, CEA-IRFU/CNRS/Universit\'e Paris Diderot, Service d'Astrophysique, CEA Saclay, F-91191 Gif sur Yvette, France}
\altaffiltext{6}{Istituto Nazionale di Fisica Nucleare, Sezione di Trieste, I-34127 Trieste, Italy}
\altaffiltext{7}{Dipartimento di Fisica, Universit\`a di Trieste, I-34127 Trieste, Italy}
\altaffiltext{8}{Istituto Nazionale di Fisica Nucleare, Sezione di Padova, I-35131 Padova, Italy}
\altaffiltext{9}{Dipartimento di Fisica e Astronomia ``G. Galilei'', Universit\`a di Padova, I-35131 Padova, Italy}
\altaffiltext{10}{NASA Goddard Space Flight Center, Greenbelt, MD 20771, USA}
\altaffiltext{11}{Department of Physics and Department of Astronomy, University of Maryland, College Park, MD 20742, USA}
\altaffiltext{12}{Istituto Nazionale di Fisica Nucleare, Sezione di Pisa, I-56127 Pisa, Italy}
\altaffiltext{13}{Istituto Nazionale di Fisica Nucleare, Sezione di Bari, I-70126 Bari, Italy}
\altaffiltext{14}{W. W. Hansen Experimental Physics Laboratory, Kavli Institute for Particle Astrophysics and Cosmology, Department of Physics and SLAC National Accelerator Laboratory, Stanford University, Stanford, CA 94305, USA}
% ...

\begin{abstract}

The detection of high-redshift ($z>$3) blazars enables the study of
the evolution of the most luminous relativistic jets over cosmic time.
 More importantly, high-redshift blazars tend to host massive black holes and can be used to constrain the space density of heavy black holes in the early Universe. Here, we report the first detection with the \fermi-Large Area Telescope of five \gm-ray emitting blazars beyond $z=3.1$, more distant than any blazars previously detected in $\gamma$-rays.
Among these five objects, NVSS J151002+570243 is now the most distant known \gm-ray emitting blazar at $z=4.31$.
These objects have steeply falling \gm-ray spectral energy distributions (SEDs) and, those that have been observed in X-rays, a very hard X-ray spectrum, both typical of powerful blazars. Their Compton dominance (ratio of the inverse Compton to synchrotron peak luminosities) is also very large ($>20$). All of these properties
place these objects among the most extreme members of the blazar population. Their optical spectra and the modeling of their optical-UV SEDs confirm that these objects harbor massive black holes ($M_{\rm BH} \sim 10^{8-10} {\rm M}_{\odot}$). We find that, at $z\approx4$, the space density of $>10^{9} {\rm M}_{\odot}$ black holes hosted in radio-loud  and radio-quiet active galactic nuclei are similar, implying that radio-loudness may play a key role in rapid black hole growth in the early Universe.

\end{abstract}


\section{Introduction} \label{sec:intro}

Blazars are the most powerful Active Galactic Nuclei (AGN) with relativistic jets oriented close to the line of sight. The jet radiation dominates their broadband emission, especially at \gm-rays  where the inverse Compton (IC) hump of the blazar spectral energy distribution (SED) peaks around tens or hundreds of MeV. The Large Area Telescope (LAT) onboard the \fermi~satellite \citep{2009ApJ...697.1071A} has already detected thousands of blazars, thus confirming that they are the most numerous population in the $\gamma$-ray sky \citep[e.g.,][]{2015ApJ...810...14A}. Nonetheless, high-redshift blazars above a redshift of 3.1 are missing  in the \fermi~catalogs, possibly due to the shift of the IC peak to lower frequencies in which LAT is less sensitive. The newly released Pass 8 photon data set \citep[][]{2013arXiv1303.3514A}, with an improved event-level analysis, substantially enhances the sensitivity of the LAT at all energies and in particular at lower energies (i.e., $<$200\,MeV). This increases the capability of the LAT to detect  spectrally soft, potentially high-$z$ blazars.

These objects typically have large bolometric luminosities ($L_{\rm bol.}>$10$^{48}$\,erg s$^{-1}$)  and host powerful relativistic jets \citep[$P_{j}\gtrsim \dot{M}c^{2}$ for a given accretion efficiency; e.g.,][]{2014Natur.515..376G}. In general, they harbor extremely massive black holes \citep[$M_{\rm BH}\sim10^{9} {\rm M}_{\odot}$;][]{2010MNRAS.405..387G}. Since blazars are highly beamed, the detection of a single blazar implies the existence of $2\Gamma^{2}$ (i.e., $\sim$400$-$600, $\Gamma$ is the bulk Lorentz factor) misaligned blazars with similar properties. Therefore, the detection of high-$z$ blazars can constrain models of supermassive black hole formation in the early Universe \citep[see, e.g.,][]{2011MNRAS.416..216V,2013MNRAS.432.2818G}. This suggests that the detection of new high-$z$ blazars will test the hypotheses of blazar evolution,  since these high redshifts constrain the time available for such extreme objects to grow.

Motivated by the recent release of the sensitive Pass~8 dataset, we perform a systematic search for new \gm-ray emitters beyond $z=3.1$ and in this Letter we report the first detection of five $z>3.1$ \gm-ray emitting blazars. Throughout, we adopt a $\Lambda$CDM cosmology with the Hubble constant $H_0=71$~km~s$^{-1}$~Mpc$^{-1}$, $\Omega_m = 0.27$, and $\Omega_\Lambda = 0.73$.

\section{Sample selection and Analysis} \label{sec:sample_red}
In order to search for high-$z$ blazars, we start from the
$\sim$1.4\,million quasars included in the Million Quasar Catalog
\citep[MQC;][]{2015PASA...32...10F}. We select all $z>3.1$ sources and
retain only radio-loud (RL) quasars with $R>10$, where $R$ is the
ratio of the rest-frame 5 GHz (extrapolated from 1.4 GHz, considering
a flat radio spectrum) to optical $B$ band flux density
\citep[][]{1989AJ.....98.1195K}, assuming an optical spectral index
$-$0.5 ($F_{\nu}\propto \nu^{\alpha}$). These 1103
objects\footnote{We exclude all the objects with a photometric
    redshift in MQC.} represent our parent sample and we analyze LAT data for all of them according to the following procedure.

For each object we use $\sim$92 months (from 2008 August 5 to 2016 April 1) of \fermi~Pass~8 source class photons and analyze them following the standard data reduction procedure\footnote{http://fermi.gsfc.nasa.gov/ssc/data/analysis/documentation/}, but with a few modifications as mentioned below. We define a region of interest (ROI) of 15$^{\circ}$ radius centered on each quasar and define a sky model that includes all $\gamma$-ray sources detected in the third \fermi-LAT catalog  \citep[3FGL;][]{2015ApJS..218...23A}, within the ROI and the isotropic and Galactic diffuse emission models \citep[][]{2016ApJS..223...26A}. The parameters of all sources within the ROI and power-law normalization factors of the diffuse models are optimized so that the sky model reproduces the data as best as possible. This is done via a binned likelihood method and we measure the significance of the detection  by means of the maximum likelihood test statistic TS=  2$\Delta \log \mathcal{L}$, where $\mathcal{L}$ represents the likelihood function, between models with and without a point source at the position of the quasar. The targets of interest are modeled with a simple power-law model leaving the prefactor and the photon index free to vary during the likelihood fitting. We consider a source to be significantly detected if TS$>25$ \citep[4.2$\sigma$;][]{1996ApJ...461..396M}.

Since we are dealing with faint sources, we modify the standard data reduction procedure as follows. We expand the energy range considered so it spans 60 MeV to 300 GeV. This allows the analysis to be more sensitive to spectrally soft $\gamma$-ray sources, i.e., blazars whose high-energy peak is shifted to lower energies ($\sim$1$-$10\,MeV) as typical for high-$z$ blazars. Moreover, Pass~8 provides an increase in the acceptance at $<$100\,MeV by up to 75\,\%, with respect to Pass 7, which translates into an improved sensitivity for  high-$z$ blazars. Because the energy resolution becomes increasingly worse at low energies we enable the energy dispersion correction in the analysis for all sources except the empirical diffuse models. A novelty introduced by Pass~8 is the characterization of the photons in PSF (point-spread function) type events,  which sub-classify the events into four quartiles by quality of angular reconstruction, with the lowest quartile (PSF0) and highest quartile (PSF3) having the worst and the best, respectively, direction reconstruction. In order to take full advantage of this improvement, we perform a component-wise data analysis for each PSF event type by considering the product of the likelihood function (sum of the logarithms) for the four PSF event types, using the SUMMED likelihood method of the Science Tools\footnote{http://fermi.gsfc.nasa.gov/ssc/data/analysis/software/}.

As in the case of our target sources, there could be faint gamma-ray emitters present in the data but not in the 3FGL catalog. Thus, we rely on an iterative procedure to discover, localize, and insert such sources into the sky model. This is done by generating a residual TS map for each ROI. The spatial positions of unmodeled excesses with TS$\geq$25 are optimized and inserted into the sky model with power-law spectra. This procedure is iterated until no significant unmodeled emission remains in the ROI.

\section{Results and Discussion}\label{sec:results}
\begin{figure*}

\hbox{\includegraphics[scale=0.6]{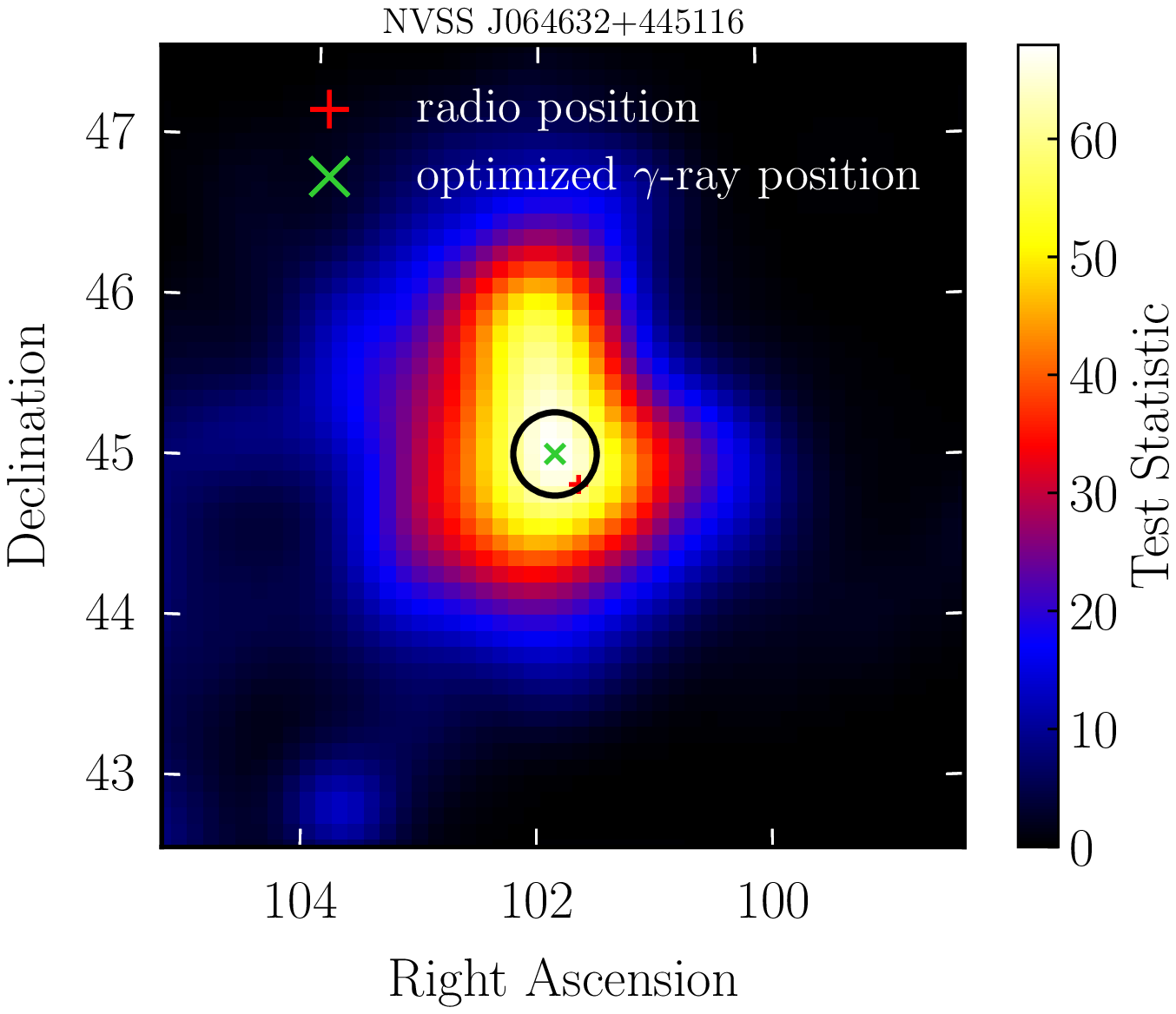}
          \includegraphics[scale=0.6]{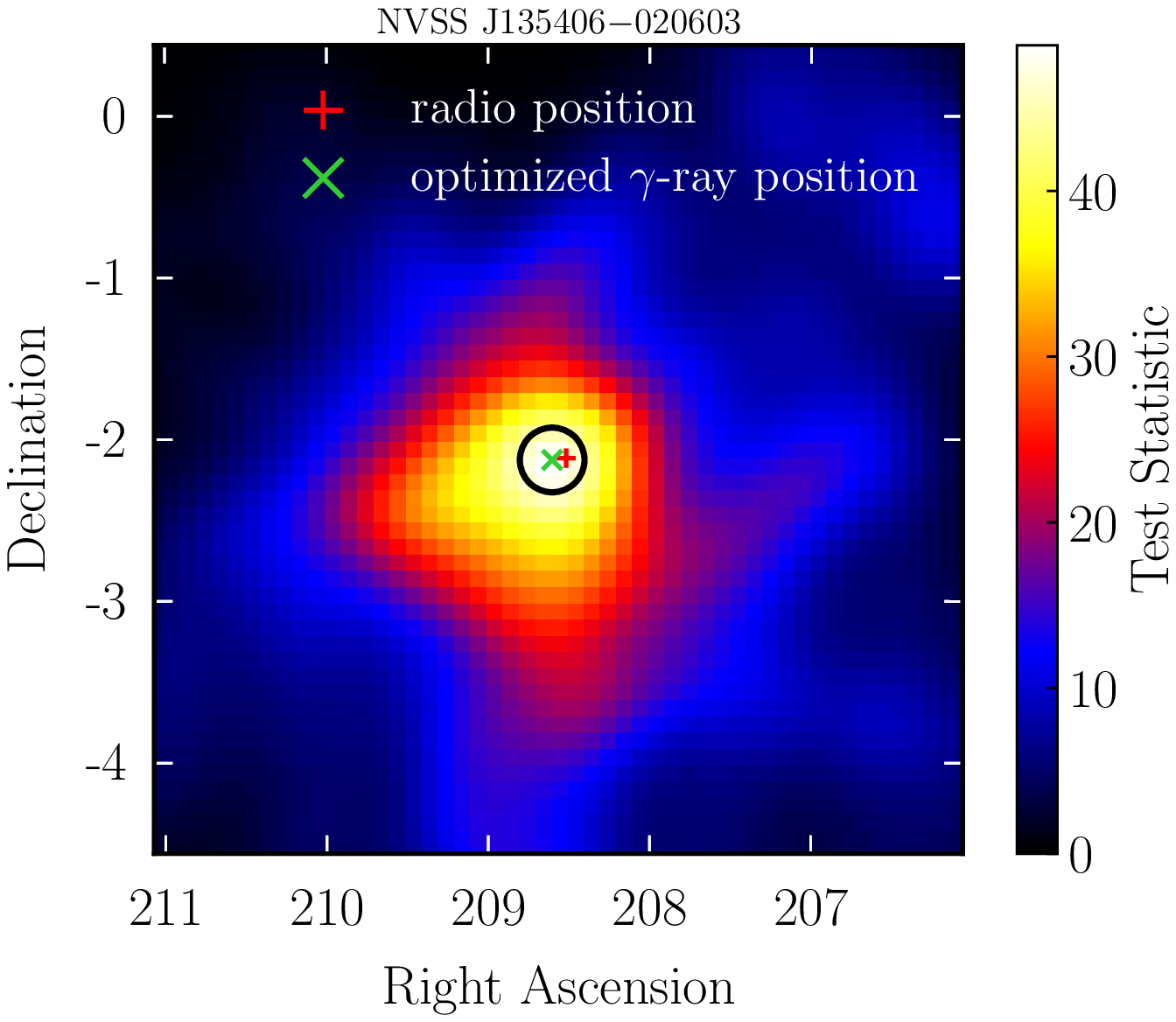}
            }
\hbox{\includegraphics[scale=0.6]{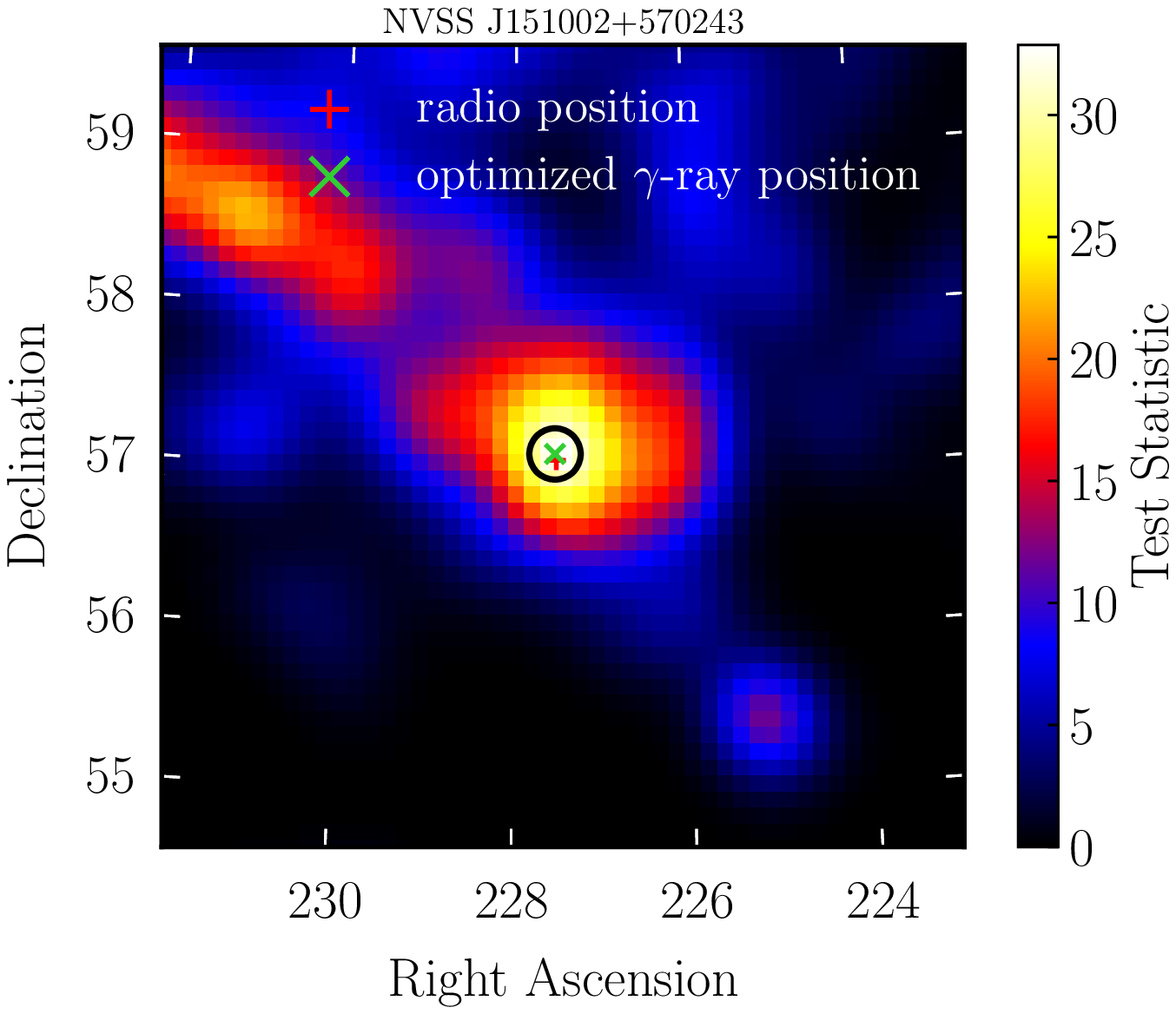}
          \includegraphics[scale=0.6]{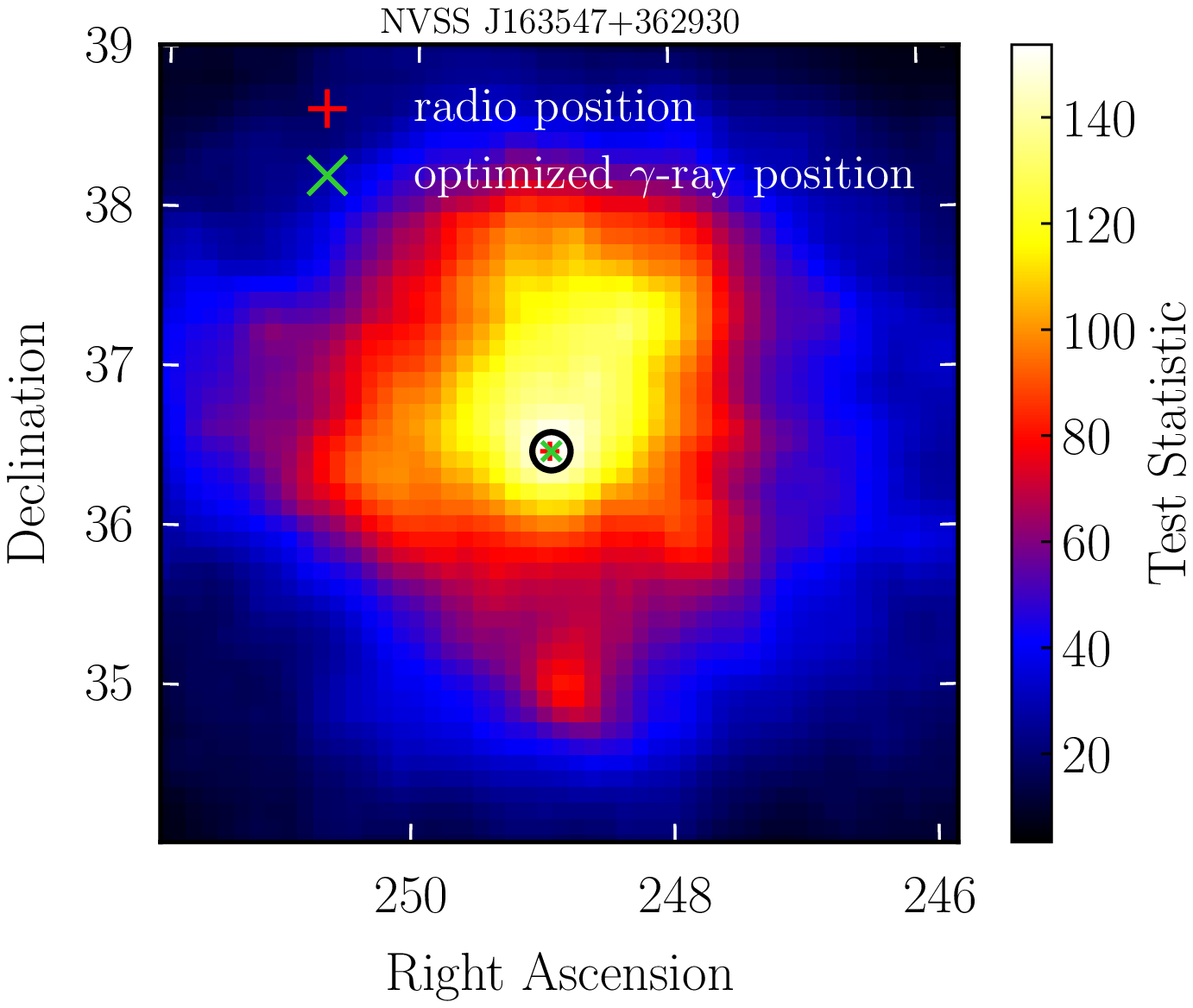}
}          
\hbox{\hspace{5.0cm} \includegraphics[scale=0.6]{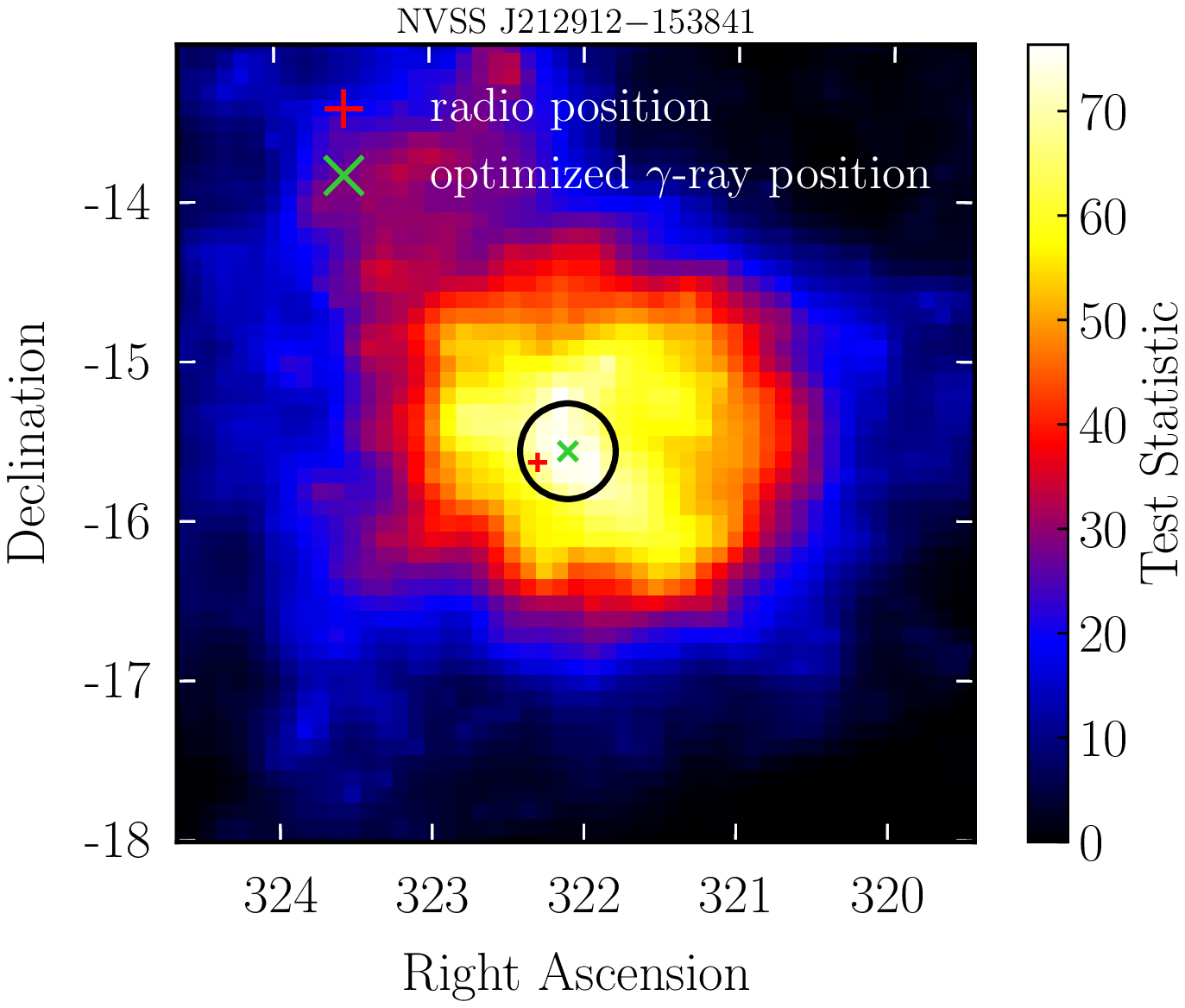}
}          
\caption{The test statistic maps of five high-$z$ quasars. The radio position (J2000), optimized \gm-ray position (J2000) and the associated 95\% error circle (in degrees) are also shown.\label{fig:tsmap}}
\end{figure*}

Our systematic search for significant \gm-ray emitters among a large
sample of RL quasars has led to the detection of five sources. A
likelihood ratio (LR) method \citep[see][for
details]{2011ApJ...743..171A} associates the detected \gm-ray sources
and their radio counterparts \citep[from NRAO VLA Sky Survey or
NVSS;][]{1998AJ....115.1693C} with high confidence (association
probability $>$80\%). By repeating the LAT analysis adopting 1000
random positions drawn from a randomized NVSS catalog\footnote{The
  NVSS catalog is randomized by mirroring the Galactic longitudes
  (glon) of the sources to 360$^{\circ}-$glon.}, we found that the
probability that any of the newly detected \gm-ray sources are
spurious is negligible. The basic information for these objects is
presented in Table \ref{tab:basic_info} where we also show the results
of the LAT data analysis. As can be seen, all the objects are
extremely RL and \gm-ray luminous quasars. According to our analysis,
{\it NVSS J151002+570243 ($z=4.31$) is now the farthest known \gm-ray
  emitting blazar}\footnote{The blazar QSO~J0906$+$6930 discovered
at $z=5.48$ by \citet{romani2006} was found to be spatially coincident with a
1.5\,$\sigma$ EGRET fluctuation, but it has not been confirmed as a $\gamma$-ray source.}.In Figure \ref{fig:tsmap}, we show residual TS maps of five detected objects, along with their radio and optimized \gm-ray positions. 

In general, high-redshift blazars are brighter at hard X-rays than in
the $\gamma$-ray band
\citep[e.g.,][]{2006AJ....132.1959R,2013ApJ...777..147S}, probably due to the shift of the
blazar SED to lower frequencies. This  could be due to the intrinsic shift
of the high-energy peak to lower energies as the bolometric
non-thermal luminosity increases \citep[][]{1998MNRAS.299..433F,2001A&A...375..739D}.
Another possible reason for the shift could be the fact that the
high-energy emission of such high-redshift blazars is powered via IC scattering off photons from the torus rather than the broad-line region (BLR), which also contributes to the lowering of the frequency of their SED peak \citep[][]{2002ApJ...577...78S}. Alternatively, the SED peaks can also shift to lower energies provided the emission region is within the dense BLR photon field. In this case, an efficient cooling of the emitting electrons will cause the lowering of the SED peaks \citep[][]{1998MNRAS.301..451G}.
 The shift of the SED causes their \gm-ray spectra to become steeper and to move slightly outside, or at the limit, of the {\it Fermi}-LAT band. Indeed, all the blazars discovered here exhibit steep \gm-ray spectra ($\Gamma_{\gamma}>2.5$, see Table \ref{tab:basic_info}). This suggests their IC peak lie at MeV energies.
 
\begin{figure*}
\gridline{\fig{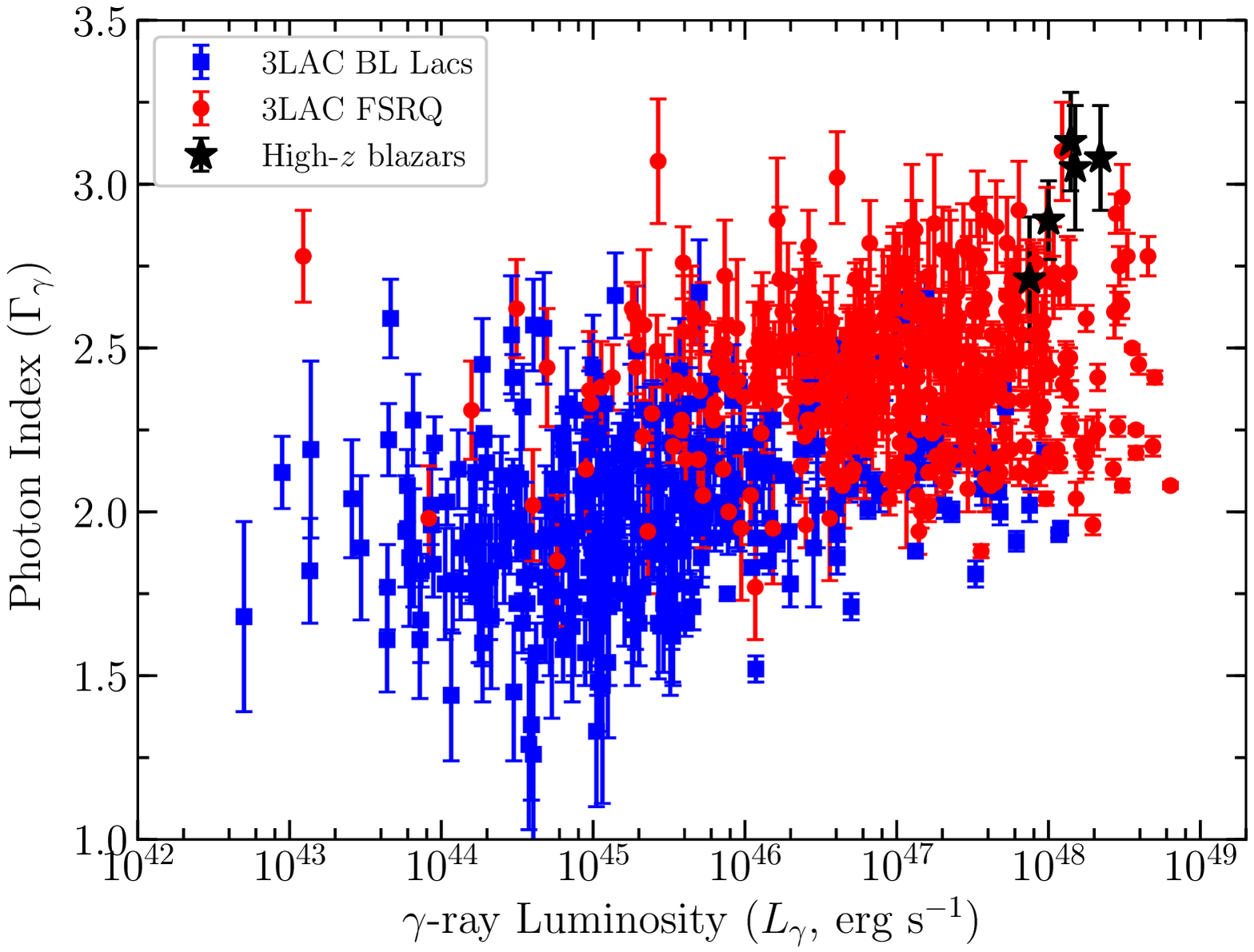}{0.5\textwidth}{(a)}
          \fig{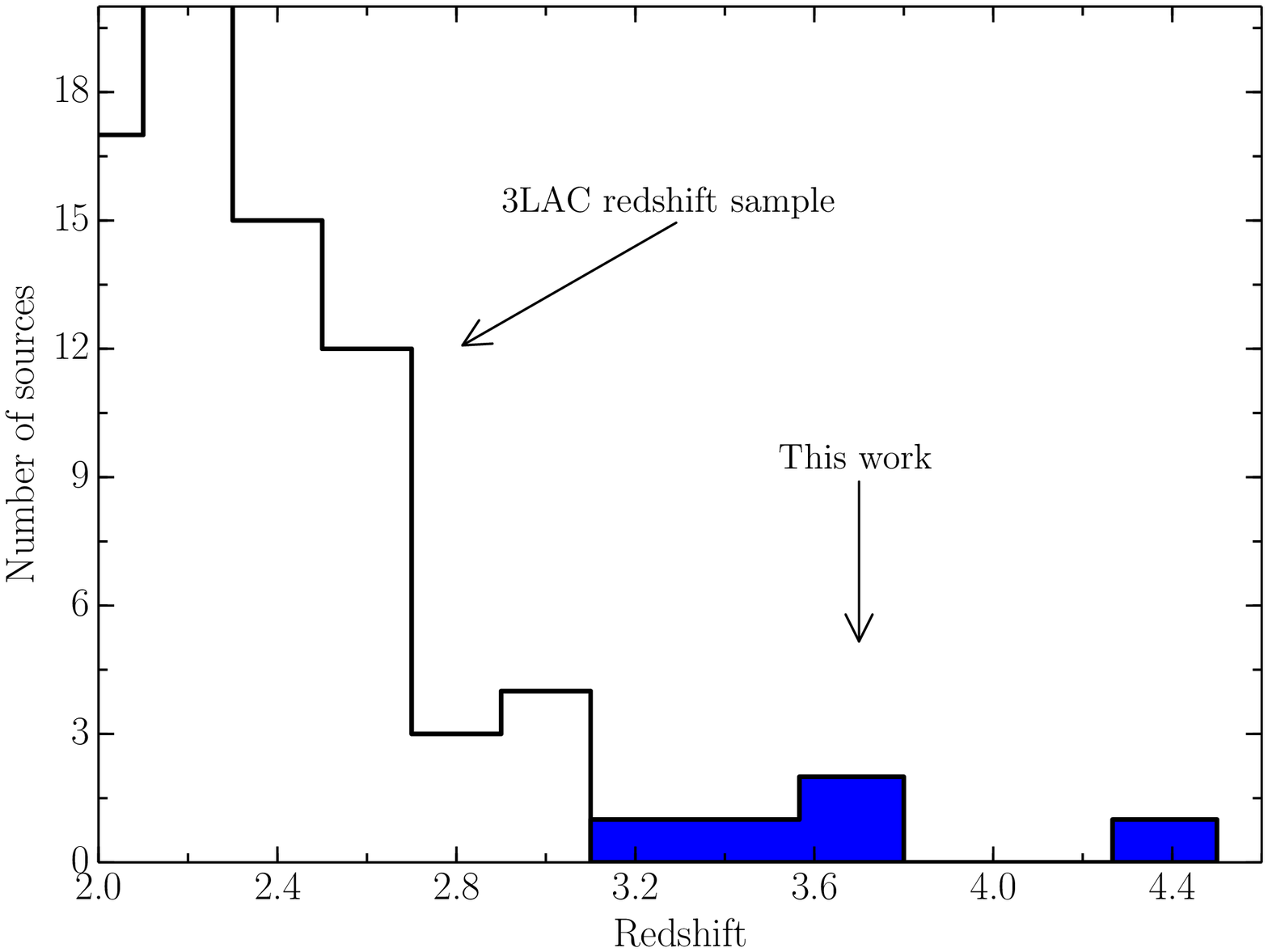}{0.5\textwidth}{(b)}
}          
\caption{Comparison of new \gm-ray detected high-$z$ blazars with 3LAC objects in, left: \gm-ray luminosity vs. photon index plane, and right: the redshift histogram. The plotted $L_{\gamma}$ and $\Gamma_{\gamma}$ are derived for the 0.1$-$300 GeV energy band, both for 3LAC and  high-$z$ blazars newly detected in $\gamma$-rays, for an equal comparison.\label{fig:3lac}}
\end{figure*}

In Figure \ref{fig:3lac}, we compare these newly detected distant objects with the blazars included in the third catalog of  \fermi-LAT detected AGN \citep[3LAC;][]{2015ApJ...810...14A}. As can be seen in the left panel, these sources occupy the region of high \gm-ray luminosities ($L_{\gamma}>10^{47}$ \lum) and soft photon indices ($\Gamma_{\gamma}> 2.5$), typical of powerful blazars.
The right panel of Figure \ref{fig:3lac} compares the redshift distributions of these newly discovered high-$z$ blazars with that of the 3LAC blazars. Though the population of distant blazars is small, this work opens a window for the study of high-$z$ blazars and may have a major impact on constraining the various physical parameters associated with the blazar population \citep[see, e.g.,][for a relevant discussion]{2016ApJ...825...74P}.

In order to understand the broadband behavior of these high-$z$ blazars we look into the literature for multi-frequency information. Though there is a paucity of such data, we found a few noteworthy observations that support the blazar nature of these objects. NVSS J064632+445116 was predicted as a candidate \gm-ray emitter by \citet[][]{2008ApJS..175...97H}, whereas NVSS J135406$-$020603 is included in the ROMA-BZCAT \citep[][]{2015Ap&SS.357...75M}. The quasar NVSS J212912$-$153841 is a hard X-ray spectrum luminous blazar and included in the 70 month \swift-Burst Alert Telescope catalog \citep[][]{2013ApJS..207...19B}. NVSS J151002+570243 is one of the best studied among all of the objects and exhibits an intense and hard X-ray spectrum \citep[e.g.,][]{1995AJ....110.1551M,1997ApJ...484L..95M,2013ApJ...763..109W}.

\begin{figure*}
\gridline{\fig{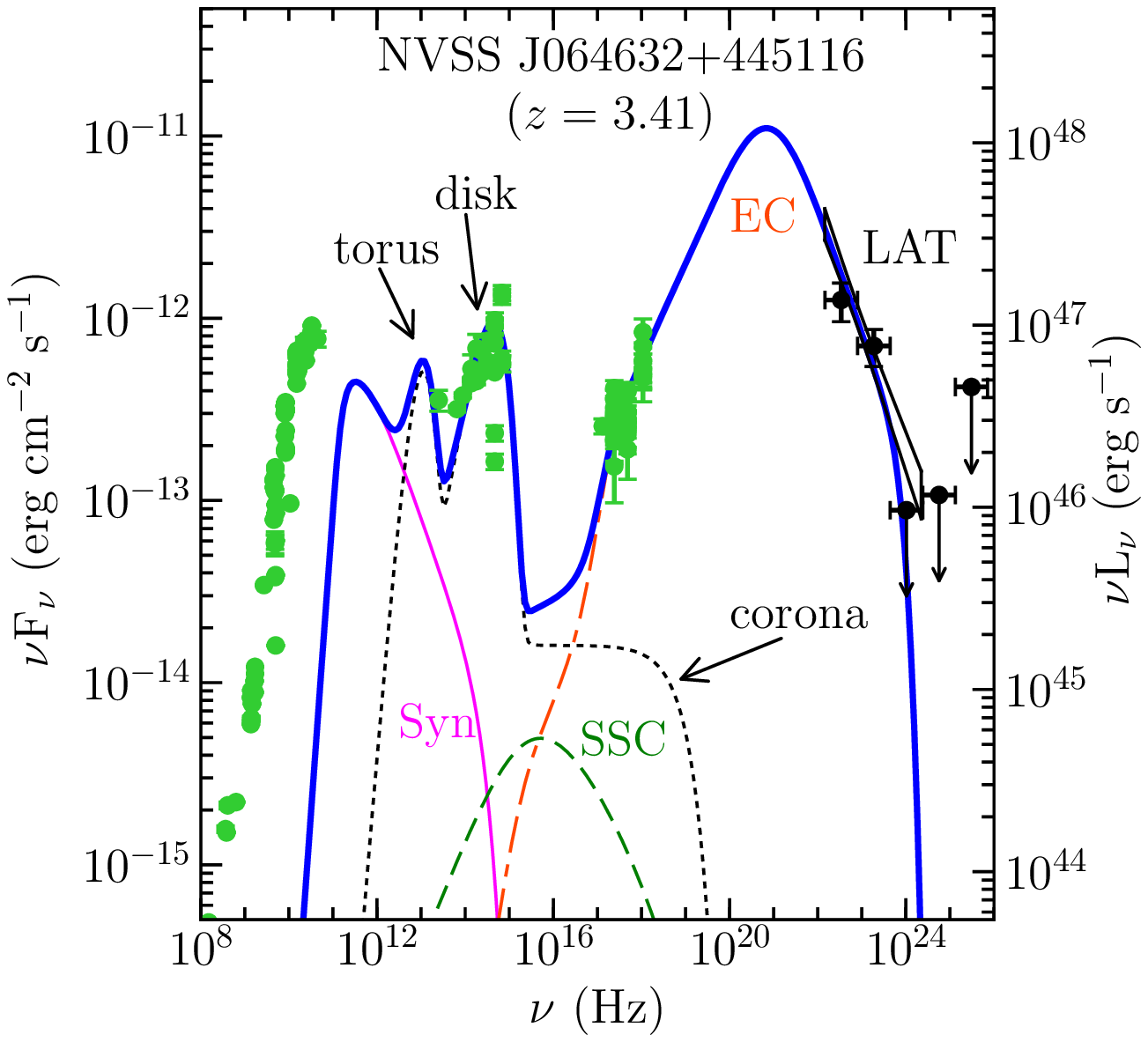}{0.5\textwidth}{(a)}
          \fig{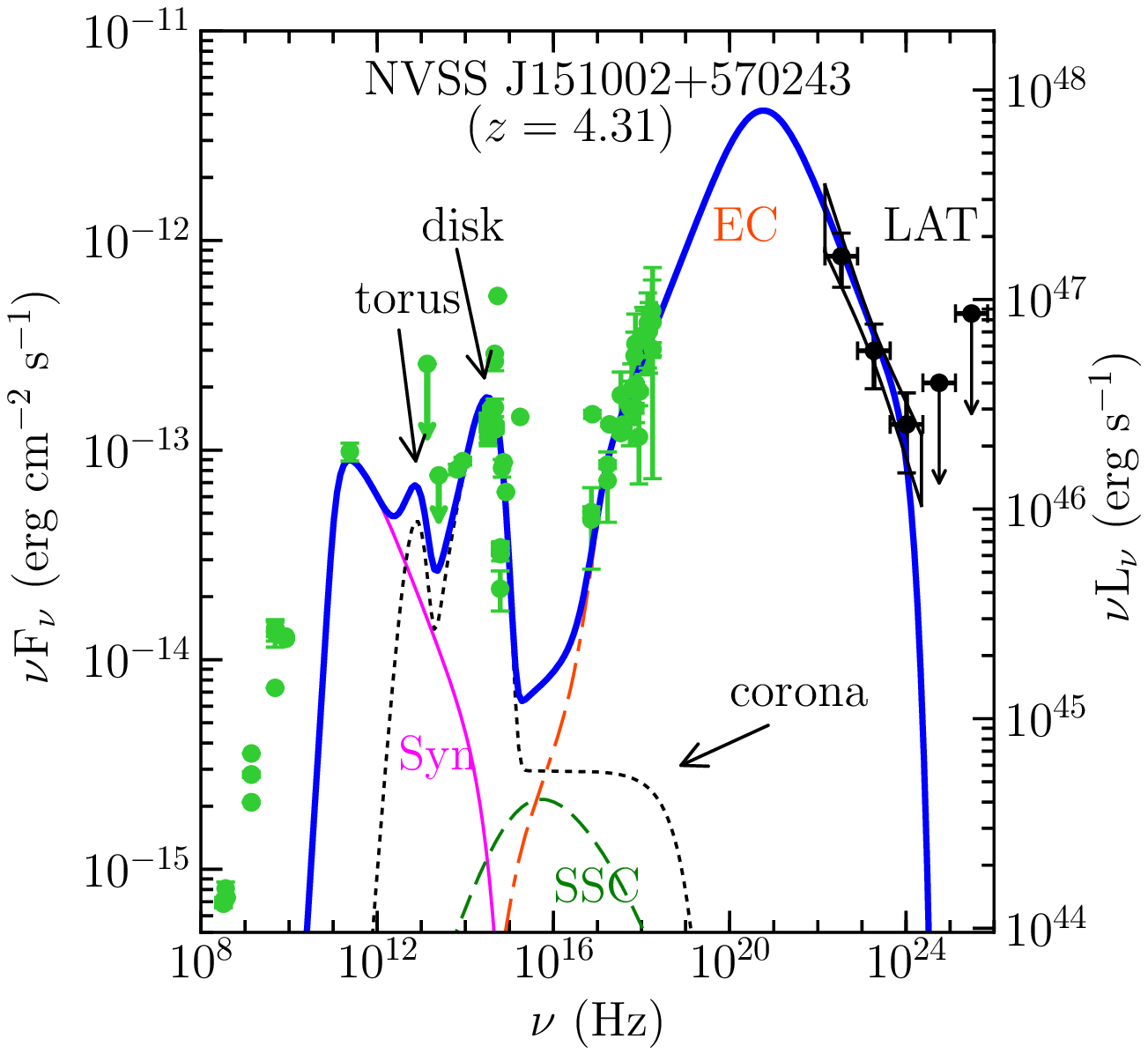}{0.5\textwidth}{(b)}
            }
\gridline{\fig{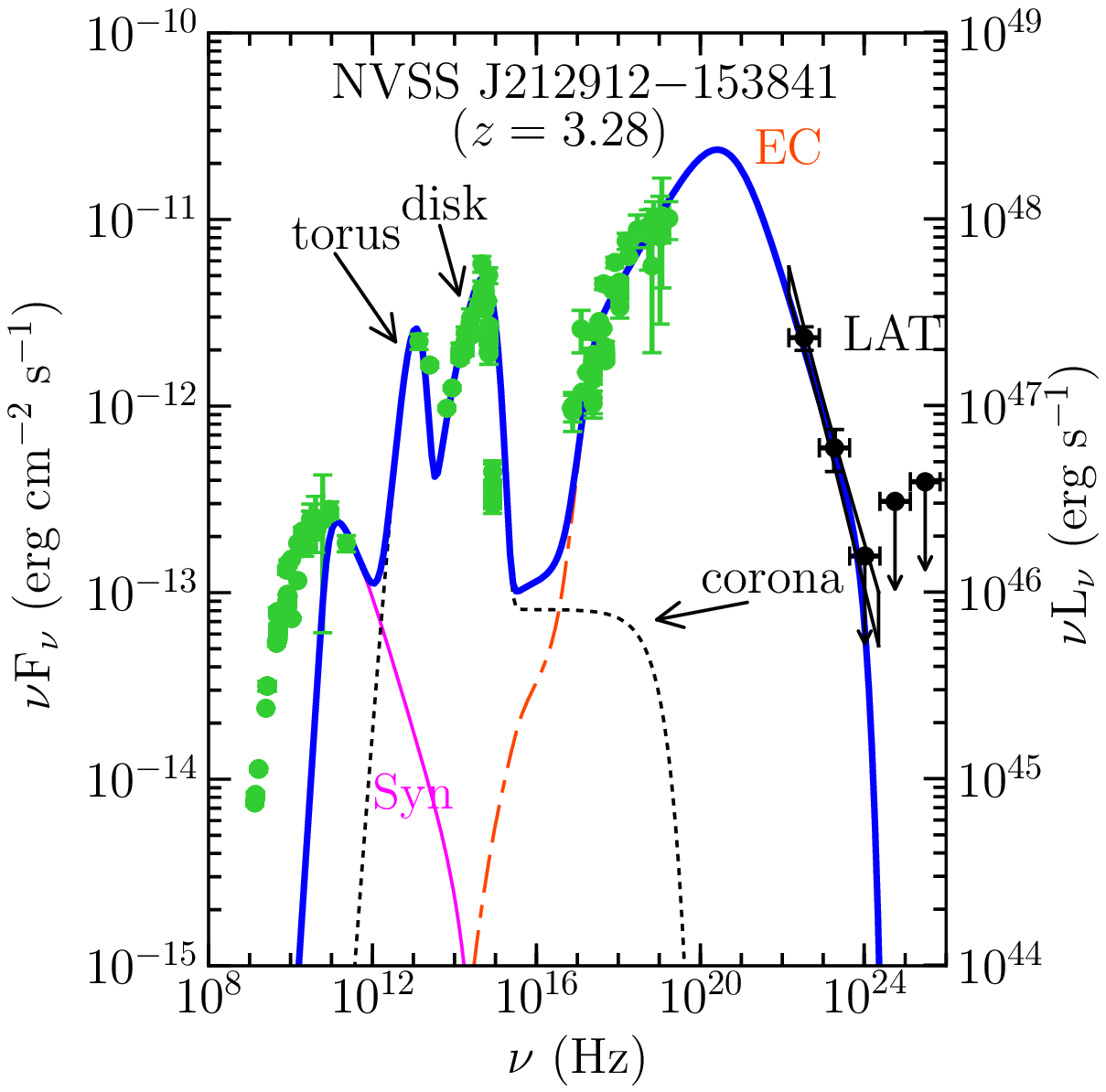}{0.5\textwidth}{(c)}
}          
\caption{The broadband SEDs of three quasars reproduced using the one zone leptonic emission model. Lime green data points are the archival observations (http://tools.asdc.asi.it/SED/) and \fermi-LAT data points and bow-tie plots are in black. The dotted black line represents thermal radiations from the IR-torus, the accretion disk, and the X-ray corona, whereas, pink thin solid, green long dashed, and orange long-dash-dash-dot lines correspond to non-thermal synchrotron, SSC, and EC emissions, respectively. The blue thick solid line denotes the sum of the contributions from all the radiative components. \label{fig:sed}}
\end{figure*}

We generate broadband SEDs of the three objects that have archival X-ray observations and model them using a simple one zone synchrotron-IC emission approach prescribed in \citet[][]{2009MNRAS.397..985G}. In brief, the model assumes a spherical emission region located at a distance $R_{\rm diss}$ from the central engine and filled with a population of highly energetic electrons that follow a broken power-law distribution. In the presence of a tangled magnetic field, the electrons lose energy via synchrotron, synchrotron self Compton (SSC), and external Compton (EC) processes. For the latter, the seed photons originate from several external AGN components: photons directly emitted from the accretion disk \citep[e.g.,][]{1993ApJ...416..458D}, from BLR \citep[][]{1994ApJ...421..153S} and from the infrared torus \citep[e.g.,][]{2000ApJ...545..107B}. The calculated jet powers and SED parameters are given in Table \ref{tab:sed_par} and the modeled SEDs are shown in Figure \ref{fig:sed}.

 In each of the three objects, the IR-UV emission is found to be dominated by an extremely luminous accretion disk ($L_{\rm disk}>10^{46}$ \lum). The X-ray spectra, on the other hand, are hard and the entire X-ray to \gm-ray band of the SED can be explained by the IC scattering off the photons originating from the BLR. A strong accretion disk radiation implies a dense BLR photon field surrounding the jet, which in turn is observable in the form of broad optical emission lines. According to our SED modeling analysis, a large BLR radiative energy density indicates that most of the high-energy emission originates from the interaction of the BLR photons with the jet electrons. This suggests that the cooling of the electrons will be efficient, and accordingly the synchrotron emission will peak at low frequencies, which is supported by the modeling results. This indicates the location of the \gm-ray emitting region to be inside the BLR. However, it should be noted that with the sparse available observations, it is not possible to tightly constrain the location of the emission region. The Compton dominance (ratio of the IC to synchrotron peak luminosities) of each of the three sources is also very large ($>20$, Table \ref{tab:sed_par}), a characteristic feature exhibited by powerful blazars. Other SED parameters are similar to those generally observed in high-$z$ blazars \citep[e.g.,][]{2010MNRAS.405..387G}. Though the data used here are mostly non-simultaneous, they indicate a {\it typical} state of the blazar rather than any period of specific activity. Also, the \gm-ray photon statistics are not good enough to search for temporal variability. Overall, the {\it Fermi}-LAT detection and the available data confirm the blazar nature of the 5 high-$z$ RL quasars.

Powerful blazars are generally found to host massive black holes at their centers \citep[e.g.,][]{2010MNRAS.405..387G,2016ApJ...825...74P}. It is, therefore, of great interest to determine the black hole mass of these \gm-ray detected quasars. In Table \ref{tab:basic_info}, we report the black hole masses for each of the five sources using information that we could find/derive from the optical spectroscopic information available in the literature \cite[][]{2012RMxAA..48....9T,2015ApJS..219...12A}. Furthermore, we also derive the masses by modeling the observed IR-UV emission for three objects (Figure \ref{fig:sed}) with a standard \citet{1973A&A....24..337S} accretion disk and the results are presented in Table \ref{tab:sed_par}. Both methods predict the existence of massive black holes ($\sim 10^{8-10} {\rm M}_{\odot}$) and match reasonably well within a factor of $\sim$three\footnote{One should keep in mind the large uncertainties ($\sim$0.4 dex) associated with the virial BH mass estimation methods \citep[see, e.g.,][]{2006ApJ...641..689V,2011ApJS..194...45S}.}, except for NVSS J151002+570243, for which the modeling approach predicts a higher black hole mass. In particular, the object NVSS J212912$-$153841 hosts one of the most massive black holes, $\sim$7 $\times 10^9$\, M$_{\odot}$, ever found in $\gamma$-ray emitting blazars, confirmed both from optical spectroscopic and disk modeling approaches.

At redshifts between 3 and 4, the space density of black holes with
$M_{\rm BH}>10^9\,{\rm M}_{\odot}$ hosted in jetted AGN is
~50\,Gpc$^{-3}$ \citep[][]{2015MNRAS.446.2483S}. This estimate is
based on the luminosity function reported in
\citet[][]{2009ApJ...699..603A}, which, at those redshifts, relies
only on five blazars. This work finds two more blazars hosting massive
black holes ($M_{\rm BH}>10^9$\,${\rm M}_{\odot}$) in the same
redshift range, considering black hole masses derived from the optical
spectroscopic information. Adopting $\Gamma=13$, derived from our SED
modeling, these two objects imply the presence of $\sim$675 (i.e.,
$2\times 2\Gamma^2$) similar systems, but with jets pointing in all
directions, in the same redshift bin. The number density related
  to these two sources can be computed as  $n=675/(V_{\rm MAX}\times f_{sky}\times
  f_{prob}\times f_{z})\approx 18$\, Gpc$^{-3}$, where $f_{sky}$=0.52 is the
  fraction of the sky covered by our parent sample, $f_z$=0.84 is the
  fraction of sources in the parent sample with a spectroscopic
  redshift, $f_{prob}$=0.66 is the fraction of $\gamma$-ray sources
  that are associated and $V_{\rm MAX}=134$\,Gpc$^{-3}$ is the
  available volume\footnote{This has been computed taking into account
  the detection efficiency of the LAT.} where
  these sources could have been detected \citep{1968ApJ...151..393S}. This brings the estimate of the space density of massive black holes hosted in jetted systems to 68$^{+36}_{-24}$\,Gpc$^{-3}$. Already at redshift 4, this implies that there is a similar number of massive black holes hosted in radio-loud and radio-quiet systems and that, given their strong evolution, above that redshift most massive black holes might be hosted in radio-loud systems \citep[][]{2010MNRAS.405..387G,2011MNRAS.416..216V}. This clearly shows that the radio-loud phase may be a key ingredient for quick black hole growth in the early Universe. To this end, the detection of high-$z$ blazars becomes very important. Currently, the most promising approaches are, (1) lowering the energy threshold of the LAT, and (2) using {\it NuSTAR}. However, the optimal instrument would be a sensitive all-sky MeV telescope, e.g., e-ASTROGAM \citep[][]{2016arXiv160803739T} and AMEGO\footnote{https://pcos.gsfc.nasa.gov/physpag/probe/AMEGO\_probe.pdf}.

\begin{splitdeluxetable*}{lcccccccBccccccc} 
\tabletypesize{\scriptsize}
\tablewidth{0pt} 
\tablecaption{Basic information and results of the \fermi-LAT data analysis of high-redshift blazars.\label{tab:basic_info}}
\tablewidth{0pt}
\tablehead{
\colhead{} & \colhead{} & \colhead{Basic information} & \colhead{} & \colhead{} & \colhead{} & \colhead{}  & \colhead{}& \multicolumn{2}{c}{} & \colhead{} & \colhead{\fermi-LAT data analysis} & \colhead{} & \colhead{} & \colhead{}\\
\colhead{Name} & \multicolumn{2}{c}{Radio position (J2000)} & \colhead{$F_{\rm 1.4~GHz}$} & \colhead{R} & \colhead{$z$} & \colhead{RL}  & \colhead{$M_{\rm BH,s}$} & \multicolumn{2}{c}{Optimized position (J2000)} & \colhead{$R_{95\%}$} & \colhead{$F_{\rm 0.06-300~GeV}$} & \colhead{$\Gamma_{\gamma}$} & \colhead{$L_{\gamma}$} & \colhead{TS}\\
\cline{2-3}
\cline{9-10}
\colhead{(NVSS)} & \colhead{hh mm ss.ss} & \colhead{dd mm ss.s} & \colhead{(mJy)} & \colhead{(mag)} & \colhead{} & \colhead{} & \colhead{$M_{\sun}$} & \colhead{hh mm ss.ss} & \colhead{dd mm ss.s} & \colhead{(degrees)} & \colhead{(10$^{-8}$ \phflux)} & \colhead{} & \colhead{(10$^{48}$ \lum)} & \colhead{}
}
\startdata
J064632+445116	  & 06 46 32.00  & +44 51 17.0    & 452 & 18.5 & 3.4 & 1253   & 9.1$^{\$}$ & 06 47 22  & +45 02 40            & 0.24 & 2.03$\pm$0.40 & 2.68$\pm$0.10 & 1.4$\pm$0.4  & 62\\
J135406$-$020603 & 13 54 06.90	& $-$02 06 03.2 & 733 & 19.2 & 3.7 & 3741 & 8.9$^{\dagger}$ & 13 54 28	& $-$02 06 43  & 0.22 & 2.32$\pm$0.52 & 2.88$\pm$0.14 & 2.5$\pm$0.8 & 44\\
J151002+570243    & 15 10 02.92  & +57 02 43.4    & 202 & 19.9 & 4.3 & 1850 & 8.5$^{\dagger}$ & 15 10 06  & +57 05 07    & 0.18 & 0.86$\pm$0.30 & 2.55$\pm$0.16 & 1.1$\pm$0.5  & 34\\
J163547+362930    & 16 35 47.24  & +36 29 30.0    & 152 & 20.6 & 3.6 & 2842 & 8.7$^{\dagger}$ & 16 35 44  & +36 29 45    & 0.13 & 5.31$\pm$0.96 & 3.15$\pm$0.14 & 7.0$\pm$0.2 & 151\\
J212912$-$153841 & 21 29 12.13  & $-$15 38 41.0 & 590 & 16.5 & 3.3 & 263  & 9.8$^{\$}$ & 21 28 25  & $-$15 34 25        & 0.33 & 2.65$\pm$0.43 & 2.82$\pm$0.10 & 1.9$\pm$0.4  & 66\\
\enddata
\tablecomments{Name, radio positions, and 1.4 GHz flux values have been adopted from the NVSS catalog. $R$ band magnitude and redshifts are taken from MQC. Radio-loudness (RL) is the ratio of the rest-frame 5 GHz (extrapolated from 1.4 GHz assuming a flat radio spectrum) to optical $B$ band flux density. $M_{\rm BH,s}$ is the logarithmic central black hole mass, in units of solar mass, derived/taken from available optical spectroscopic information: $^{\$}$\citet[][]{2012RMxAA..48....9T}, $^{\dagger}$\citet[][]{2015ApJS..219...12A}. $R_{95\%}$ is the 95\% error radius derived from the analysis. The \gm-ray flux and apparent luminosity are in the energy range of 0.06$-$300 GeV.}
\end{splitdeluxetable*}

\begin{table*}
\begin{center}
\caption{Summary of the parameters used/derived from the modeling of the SED of the objects shown in Figure~\ref{fig:sed}. The viewing angle is taken as 3$^{\circ}$ for all of them.}\label{tab:sed_par}
\begin{tabular}{lccccc}
\tableline
Parameter 																										   & J0646+4451 & J1510+5702 & J2129$-$1538 \\
\tableline
\tableline
Slope of the electron energy distribution before break energy ($p$)               & 1.8           & 1.8             & 2.2    \\
Slope of the electron energy distribution after break energy ($q$)                 & 4.4           & 4.1             & 4.5     \\
Magnetic field in Gauss ($B$)                                                                       & 2.1           & 1.4             & 1.3     \\
Particle energy density in erg cm$^{-3}$ ($U'_{e}$)                                     & 0.009       & 0.029         & 0.002 \\ 
Bulk Lorentz factor ($\Gamma$)                                                                  & 12            & 11              & 14      \\
Minimum Lorentz factor ($\gamma'_{\rm min}$)                                          & 1              & 1                & 1        \\
Break Lorentz factor ($\gamma'_{\rm b}$)                                                  & 72          & 82              & 51      \\
Maximum Lorentz factor ($\gamma'_{\rm max}$)                                         & 2e3          & 3e3            & 2e3    \\
Size of the emission region in parsec ($R_{\rm blob}$)                                  & 0.025       & 0.017         & 0.059 \\
Dissipation distance in parsec ($R_{\rm diss}$)                                             & 0.25         & 0.17           & 0.59 \\
Size of the BLR in parsec ($R_{\rm BLR}$)                                                    & 0.37         & 0.22            & 0.79 \\
Black hole mass in log scale, in units of solar mass  ($M_{\rm BH,m}$)           & 9.60          & 9.48           & 9.85   \\
Accretion disk luminosity in log scale ($L_{\rm disk}$, erg s$^{-1}$)             & 47.11        & 46.65         & 47.78 \\
Accretion disk luminosity in Eddington units ($L_{\rm disk}/L_{\rm Edd}$)    & 0.26          & 0.12           & 0.68   \\
Fraction of the disk luminosity reprocessed by the BLR ($f_{\rm BLR}$)         & 0.1            & 0.1              & 0.1      \\
 Fraction of the disk luminosity reprocessed by the IR-torus ($f_{\rm IR}$)     & 0.5            & 0.5              & 0.5      \\ 
\hline
Compton dominance (CD)																				  & 25				& 47               & 99 \\
Jet power in electrons in log scale ($P_{\rm e}$, erg s$^{-1}$)                     & 44.88         & 44.94        & 45.14 \\
Jet power in magnetic field in log scale ($P_{\rm B}$), erg s$^{-1}$              & 46.15         & 45.38        & 46.58 \\
Radiative jet power in log scale ($P_{\rm r}$, erg s$^{-1}$)                          & 45.89         & 45.70        & 46.31 \\
Jet power in protons in log scale ($P_{\rm p}$, erg s$^{-1}$)                        & 47.38        & 47.45         & 47.89 \\
\tableline
\end{tabular}
\end{center}
\end{table*}

\acknowledgments
We are grateful to the referee for insightful comments. The \textit{Fermi}-LAT Collaboration acknowledges support for LAT development, operation and data analysis from NASA and DOE (United States), CEA/Irfu and IN2P3/CNRS (France), ASI and INFN (Italy), MEXT, KEK, and JAXA (Japan), and the K.A.~Wallenberg Foundation, the Swedish Research Council and the National Space Board (Sweden). Science analysis support in the operations phase from INAF (Italy) and CNES (France) is also gratefully acknowledged. Part of this work is based on archival data, software, or online services provided by the ASI Science Data Center (ASDC).
\vspace{5mm}
\facilities{\fermi-LAT}

\end{document}